# Demonstration Paper: Wirelessly Sensing Medication Administration

Cyber-Physical Event Detection and Notification Utilizing Multi-Element Chipless RFID

Xiaofu Ma
xfma@vt.edu

Thaddeus Czauski
czauski@vt.edu

Taeyoung Yang
mindlink@vt.edu

Jeffrey H. Reed
reedjh@vt.edu

Wireless@Virginia Tech, Department of Electrical and Computer Engineering
Virginia Tech
Blacksburg, Virginia 24061, USA

## ABSTRACT
Medication administration is one pathway by which Adverse Drug Events (ADE) can occur. While Electronic Medical Administration Record (eMAR) systems help reduce the number of ADEs, current eMAR implementations suffer from workarounds that defeat safety and verification mechanisms meant to limit the number of potential ADEs that occur during medication administration. In this paper, we introduce Multi-Element ChipLess (MECL) RFID tags which enable real-time event notifications through event signatures. Event signatures correspond to the physical configuration of different RFID elements in a chipless RFID tag. Augmenting physical objects, such as a pill container, with MECL-RFID can allow caregivers to detect the moment a particular pill container is opened or closed. We present the fundamentals behind real-time event detection using MECL-RFID and propose a cyber-physical intervention system that can be used to reduce ADEs through real-time event monitoring and notifications sent to clinicians administering medication. We also present a prototype MECL-RFID to demonstrate potential future improvements to eMAR systems that minimize ADEs.

## 1. INTRODUCTION
Adverse Drug Events (ADE) can negatively affect patients in many scenarios, including hospitals where clinicians administer medication to patients, or home care where patients self-administer prescribed medication. Investigating the aforementioned medication administration scenarios, studies on medication administration have concluded that ADEs have the potential to be avoided [1, 2]. To reduce the number of ADEs, Electronic Medical Administration Record (eMAR) systems are being adopted. eMAR systems such as BarCode Medicine Administration (BCMA) have demonstrated their effectiveness in reducing ADEs, and Poon et al. [3] found that in some cases BCMA has reduced the potential for ADEs by 51%, illustrating the exciting potential that eMAR systems can have on mitigating ADEs.

While eMAR systems have improved medication administration safety in hospitals, there still exist challenges to ensuring potential ADEs during medication administration are minimized. The advantages of BCMA systems are counterbalanced by workarounds that can limit some of the safety benefits of BCMA [4]. Another issue is a perceived lack of eMAR system usefulness to support patient care by clinicians which could reduce eMAR acceptance [5]. Furthermore, a growing challenge is minimizing emergency hospitalizations for older Americans due to ADEs when self-administering medication at home [2].

If medicine bottles could automatically detect when they were opened as well as when they were closed, then eMAR systems might be considered more useful in hospital and home care scenarios for detecting drug events and attempting to mitigate ADEs in real-time at the point of administration. To achieve real-time interventions, we propose a Multi-Element ChipLess Radio-Frequency IDentification (MECL-RFID) method that can be integrated into medicine bottles and used to detect and record drug events, and notify clinicians or patients at the point at which the medication is about to be administered of potential ADEs. In this paper, we propose using cyber-physical systems (CPS) and cognitive radios, for implementing Health Information Technology (HIT) systems, such as eMAR where drug events (e.g. opening a medicine container and closing the container) are wirelessly sensed by advanced cognitive wireless sensors. Event detection is then automatically recorded and checked for errors, and cyber-physical interventions may be performed where clinicians are automatically notified of warnings and errors via mobile devices (e.g.

smartphones, tablets) or on larger display devices such as televisions:

- eMAR applications pose unique monitoring and intervention challenges for improving patient outcomes, and new issues arise for preventing potential ADEs when patients follow a prescription regimen in their homes. Additionally, automated sensing of ADEs requires minimizing wireless sensing complexity to enable real-time ADE detection and intervention. These issues are described in Section 2, and Section 3 provides an overview of related works in similar research areas.
- In order to provide automated detection of events, we propose a novel MECL-RFID system. The unique physical configuration of the multiple RFID elements is leveraged to enable intelligent event detection, where the spatial configuration of the multiple chipless RFID elements provides event notifications. The MECL-RFID tag and the operating principles behind intelligent event detection are elaborated in Sections 4.1 thru 4.3.
- Section 5 provides an evaluation of different algorithms for processing event detection using MECL-RFID.
- In Section 6, we present a system level discussion of integrating MECL-RFID into a cyber-physical eMAR and our proof of concept research prototype implementation of the MECL-RFID augmented medication bottle. We demonstrate that our chosen event detection algorithm (Section 4.3.2) is able to perform real-time interventions, and we share some experiences using mobile devices and MECL-RFID to enable cyber-physical interventions. We conclude our discussion in Section 7.

## 2. MOTIVATION AND CHALLENGES

Ensuring the patient's '5 rights' (i.e. right drug, right dose, right patient, right form, right route) is essential to ensuring excellent care for patients when administering any form of medication. In this section, we outline the motivating factors behind our work, mainly the importance of minimizing potential ADE complications that can arise when administering medication in healthcare settings or home-care settings. We also identify eMAR challenges that our research aims to address including: (*i*) The difficulties of needing to manually scan medication to identify drug events, and (*ii*) wireless sensing complexity limitations in chipless RFID technology that makes automated and wireless event detection an arduous task.

### 2.1 Automated ADE Detection is Typically Limited to Health Care Provider Facilities

A growing concern is outside the hospital where eMAR systems may not be available. With increasing lifespans, people are living longer and there is an increased demand for physicians specializing in the field of geriatric medicine. Furthermore, many older patients' (aged 65 or older) goals may be to remain independent and take care of themselves by visiting a primary care physician and then following a prescribed home medication regimen. In many scenarios where older patients are self-administering at home, 40% may be managing upwards of 5 different prescriptions [2]. Managing many different home medication regimens may be difficult, and errors in dosing may contribute to ADEs. Budnitz et al. [2] found that nearly 66% of hospitalizations for older patients were due to unintentional medication overdoses. Many eMAR systems are only deployed in dedicated healthcare environments (e.g. hospitals). It is important to consider that eMAR systems and their ADE reduction benefits are not universally available in all healthcare settings due to the cost and effort required to implement and maintain eMAR systems [1]. An open research challenge is designing eMAR systems which could be easily deployed in multiple environments to bring the ADE reduction benefits of eMAR into more areas of healthcare both inside and outside healthcare facilities.

### 2.2 ADE Event Detection with Static Markers is Cumbersome and Intrusive

BCMA and Integrated Circuit (IC) based RFID eMAR implementations typically rely upon close physical proximity to the patient and prescription in order to scan and verify static information encoded on markers (i.e. barcodes or IC-RFID tags). This requires clinicians to add additional steps to their workflow and manually scan each medication to be administered as well as the patient's marker before administering medication. Scanning each item is an intrusive process, and in the interest of efficiency some administering clinicians may introduce workarounds such as scanning one medication packet multiple times if the packets contain the same medication and dose [4].

The multi-scan workaround can potentially subvert the eMAR and introduce ADEs since many medication packets look alike; if an administering clinician scans the same packet multiple times, but different packets contain different doses then the variation in dosing will not be screened by the eMAR and a potential ADE is introduced. This is but one example, Koppel et al. [4] outline additional technology, process, organizational, patient and environment related workarounds in BCMA eMAR systems that could hinder the eMAR's effectiveness. Furthermore, Holden et al. [5] suggest that the cumbersome nature of conventional eMAR systems can be attributed to some administering clinicians' perceptions that BCMA eMAR systems are of minimal value in streamlining patient care, and is one contributing factor to workarounds that result in suboptimal use of BCMA. The rise of workarounds applied to eMAR systems is another open research challenge, where eMAR usability must be altered and enhanced to complement clinicians' workflows to ensure that the ADE reduction benefits of eMAR can be achieved while not impacting or intruding upon clinicians' and caregivers' workflows when providing care to patients.

### 2.3 Chipless RFID Decoding is Complex

Recent studies on chipless RFID tags [6-13] are based on electromagnetic scattering theory. The natural resonances of chipless tags can be excited by a stimulus signal (i.e. a signal sent from another device or base station), and the chipless tag's information is encoded within the backscattered signal produced by the metallic pattern of the tag.

In much of the research to date, an expensive and complex vector network analyzer was used for collecting scattered-field data (i.e. radio frequency signatures) from chipless RFIDs. Post processing for detection and identification of the RFID data was carried out after signal data was captured on commercial computing hardware. Because of the high-sampling rate and complexity of the vector network analyzer required to analyze chipless RFID signals, implementing a computationally inexpensive chipless RFID reader is difficult. Alternate chipless RFID decoding methods are needed to provide real-time event detection in resource constrained environments, in contrast with current post-event store and process methods.

## 3. Related Works

Chipless RFID and CPS are concepts that are well known, and this work is not the first to propose applying chipless tag technology or CPS for improving eMAR systems. In this section we provide a brief overview of previous work in the field which has inspired our work, and elaborate on the concepts this paper contributes to the field.

### 3.1 Chipless RFID

There are various different types of chipless RFID being studied. Preradovic and Karmakar [23] introduce a taxonomy of chipless RFID operating theories which are centered around (*i*) Time-Domain Reflectometry chipless RFID, (*ii*) Spectral Signature based chipless RFID, and (*iii*) Amplitude/Phase Modulation based chipless RFID. MECL-RFID would correspond to the spectral signature category as the frequency notches produced are equivalent to, "the presence or absence of a resonant peak at a predetermined frequency" feature that defines the spectral signature category [23]. MECL-RFID is influenced by the work of [6-13], where the physical properties of a chipless RFID antenna correspond with the signature or frequency notches produced by the antenna. MECL-RFID extends the initial frequency notch concept by enabling the notch locations to be dynamic based on the physical interplay of the chipless RFID elements which compose a MECL-RFID tag.

### 3.2 Cyber-Physical Systems

CPS generally illustrate three properties: (*i*) the ability to sense the physical world, (*ii*) the ability to interpret the physical sensor data via a computation in the cyber domain, and (*iii*) the ability to act upon the result of the digital computation and influence the physical domain. MECL-RFID by itself is one important piece of a larger system. To enable the real-time sensing and intervention capabilities of our prototype, we leveraged and built upon several of the general CPS concepts outlined above. A complete discussion of cyber-physical concepts is outside the scope of this paper, but a comprehensive overview of CPS and example applications are described by Hu [24].

### 3.3 Clinical Applications of RFID

Previous work in medication administration predominantly focuses on the usage of static information stored on an IC-RFID tag. For example, Wu et al. [25] use IC-RFID tags to identify a patient's identity and compare the drugs being administered to the patient. Becker et al. [26] not only use IC-RFID tags to monitor medication administration, but also introduce a 'smart drawer' that can detect when the drawer is opened or closed based on the ambient light detected by a sensor in the drawer as well as the drawer's direction of acceleration.

In all these cases the static RFID information needs be manually scanned at close proximity. While Becker et al. are able to detect other events (i.e. opening/closing of the medicine drawer), they rely upon additional sensors (i.e. ambient light sensors and accelerometers). The manual nature of IC-RFID and barcodes limit event detection and does not necessarily improve or simplify medication administration workflow. Furthermore, MECL-RFID stands out among traditional clinical applications of RFID as MECL-RFID can dynamically respond to changes in the environment.

## 4. WIRELESS EVENT DETECTION AND CYBER-PHYSICAL INTERVENTION

From Section 1, there are many benefits when eMAR systems can automatically detect drug events and take the appropriate action just in time to prevent an ADE. In this section, we present the design details and operational theory behind for our proposed cyber-physical intervention system, and address the challenges outlined in Section 2. The main elements that power cyber-physical intervention for detecting and mitigating potential ADEs are centered around a cognitive radio network that can detect drug events in real time by sensing MECL-RFID tags. These tags provide feedback on medicine container status (i.e. open, closed). The cognitive radio network can then observe and report events to medical applications like an eMAR, and applications can send alerts to clinicians or patients if potential ADEs are detected when preparing to administer medication or simply record that the administration event took place if no ADEs are detected.

### 4.1 Cognitive Sensing within Wireless Healthcare Environments

Current commercial RFID tag readers are often designed for a fixed frequency band with no adaptability for other applications (i.e. the tag readers are only compatible with one type of tag on a particular frequency). Conventional tag readers are neither designed to utilize other wireless spectrum nor the newly released spectrum for healthcare use on a secondary basis [14]. To exploit the available spectrum for medical use, provide medical event detection services, and to increase the flexibility of the sensing platform to address unknown future needs, we base the cognitive sensing architecture around flexible software-defined radio (SDR) and cognitive radio (CR) techniques [15, 16]. SDR and CRs are capable of adapting radio operations to optimize the system performance in many different environments. SDR and CR enable dynamic access to the spectrum for various RFID technologies as well as other wireless applications. A SDR-based wireless CR network allows designers to reconfigure the system for use with future RFID tags by simply updating software. As part of a cyber-physical intervention system, SDR and CR techniques are used for communicating with RFID tags and monitoring for changes to event signatures, which are described in the next section.

### 4.2 Multi-Element ChipLess (MECL) RFID

Detecting events wirelessly requires a marker to dynamically respond to, or change with, a corresponding change in the marker's environment. As discussed in Section 2.2, barcodes and IC-RFID tags store static information. MECL-RFID aims to enable RFID tags to report dynamic information, based on the tag's environment. Specifically, several chipless RFID elements with different electromagnetic properties are composed into a single MECL-RFID chip, where the interplay of the different chipless RFID elements enables the MECL-RFID tag to dynamically change its status information in response to environmental changes.

The repositioning of a MECL-RFID augmented object's shape or form will change the overall scattering properties of the constituent chipless RFID tags and the physical reconfiguration of the MECL-RFID creates a unique event signature that is observable by the CR network. An example of detecting different events using MECL-RFID tags in the augmented bottle is illustrated in Fig. 1.

The chipless RFID tag structure is printed on the surface of a container which allows 3-bits of information to be stored on the tag. If extended code information is used, the MECL-RFID can include patient name, medication name, dosage, expiration date, prescribing physician, and other information commonly associated with the label on the medicine. The detected scattered electric fields from the tags produce frequency notches when scanned as shown in Fig. 1a. The location of the frequency notches represents coded identification and any additional coded information.

In addition to the tags on the body of the container, a distinctive tag structure is also printed on the lid of the container. When the lid is attached to the container (Fig. 1b), the event signature, or frequency notches, produced by the MECL-RFID change as the frequency notch locations are shifted to a lower frequency as well as overall narrowing of the individual frequency notches. The shifting-frequency phenomenon is due to the loading effect of the lid tag structure to the tag within the body of the container. The overall frequency-shifting phenomenon with a reduced notch width can be used to create differing event signatures to automatically record events (e.g. bottle opened, bottle closed) without the need for the additional step of manually scanning the medicine to be administered as described in Section 2.2.

We combine the CR network with the event signatures generated by MECL-RFID tags, and cyber-physical intervention becomes possible in almost any environment where the physical state of the bottle is evaluated by an eMAR to determine if the '5 rights' are being properly adhered to and notifications are pushed to clinicians and patients regarding important ADEs (e.g. incorrect dose). Integrating MECL-RFID and CPSs addresses the challenges described in Section 2.1 and provides a pathway to address current eMAR deployment workarounds (Section 2.2) using real-time event signature detection with a CR basestation or network of CR radios.

Although there are several chipless RFID technologies being researched (Section 3.1), our choice of utilizing a metallic tag design was influenced by several factors: (*i*) the metallic tag structure on the lid does not need to physically connect with the chipless RFID tags on the medical container. Both chipless RFID structures on the lid and the container could be electrically (capacitively) or magnetically (inductively) coupled through a physical gap. Therefore, the tag structures are not damaged by frequent opening and closing of the containers; (*ii*) metallic chipless RFID technology can be easily applied through additive manufacturing techniques; and (*iii*) the backscattering produced from the excitation signal by the metallic RFID structure would be stronger than other materials (e.g. chemical or magnetic structures), which provides a larger detection range.`

## 4.3 Chipless RFID Detection Methods

Existing chipless RFID detection approaches are based on the singularity-expansion method (SEM) model [17]. The basic idea of SEM is that a chipless RFID's time-domain scattered field response to electromagnetic stimuli can be represented by a sum of damped exponential terms, shown in Equation (1).

$$E_{scattered}(t) = \sum_i R_i e^{-(\alpha_i + j\omega_i)t} + n(t) \qquad (1)$$

In Equation (1), $R_i$ is residue, $\alpha_i$ is damping factor, and $\omega_i$ is angular frequency of each complex pole ($s_i = \alpha_i + j\omega_i$). This model comes from practical observations that electromagnetic responses to impulsive excitations generally take the form of a summation of damped sinusoids. Overall noise, including RF environment and/or numerical noise, is represented as $n(t)$ in Equation (1). For an exact representation of the scattered field data, an infinite sum of complex exponentials is required. In practice, however, the response can be modeled accurately using only a few dominant complex poles due to the band-limited nature of typical signals.

### 4.3.1 Matrix Pencil Method (MPM):

A popular pole detection approach used for chipless RFID is based on the matrix pencil method (MPM) due to MPM's superior detection capability in noisy RF environments [18]. MPM exploits the structure of the matrix pencil of the underlying signal, and singular-value decomposition (SVD) is applied to extract the signal from the noisy sequence.

However, the matrix pencil method typically requires complex time-domain analysis combined with a high sampling rate. Alternatively, MPM requires both amplitude and phase information in the frequency domain. In lab environments, the typical time resolution required to use MPM for decoding chipless RFID information is less than 0.1 ns [6, 19]. Thus, the sweeping frequency range of the chipless RFID reader needs to be greater than 10 GHz. As noted in Section 2.3, the MPM detection approach requires computationally powerful and complex sensing equipment, making MPM difficult to implement.

MPM detection performance may be characterized using a normalized decoding error between the exact complex pole ($s_i$) and the estimated complex pole ($\hat{s}_i$) for a given signal-to-noise ratio (SNR), which can be defined as shown in Equation (2).

$$MPM_{Err}(SNR) = \sum_i \|s_i(SNR) - \hat{s}_i(SNR)\| \qquad (2)$$

### 4.3.2 Pattern Recognition Approach (PRA):

An alternate detection approach for chipless RFID can be based on similarity measures. Our pattern recognition approach (PRA) relies on preprocessing of the data to represent patterns from the feature space [20]. In order to detect and recognize the encoded binary information within the chipless RFID tag, the pattern matrix can be defined as:

$$x = [[f_{notch}] \quad [W_{notch}] \quad [D_{notch}]] \quad (3)$$

where $f_{notch}$, $W_{notch}$, and $D_{notch}$ are the spectral locations, widths, and depths of the frequency notches, respectively. Euclidean distance is adopted to evaluate the degree of dissimilarity between the defined similarity template and the decoded data being evaluated for the existence of a known pattern. The defined patterns do not require phase information of the scattered field data, which allows PRA decoding of chipless RFID to be implemented in a cost-effective manner, addressing the computational complexity challenges (Section 2.3).

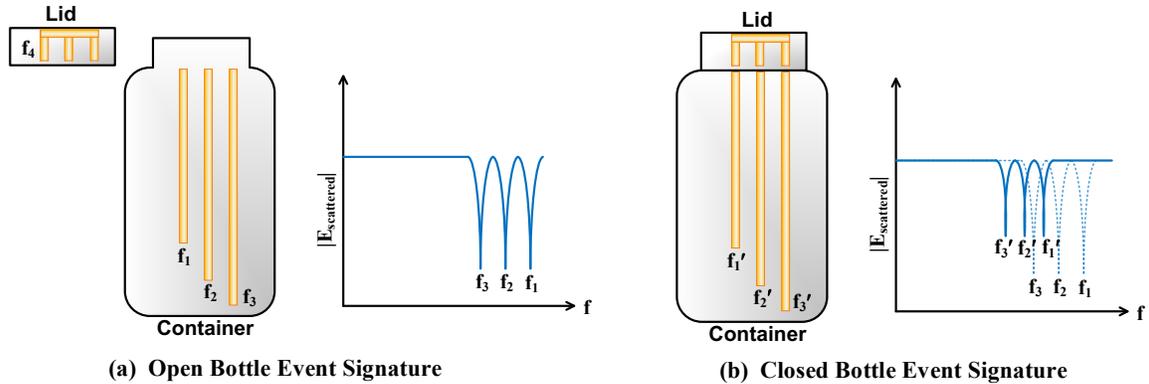

(a) Open Bottle Event Signature  (b) Closed Bottle Event Signature

**Fig.1: Using MECL-RFID Event Signature Detection: (a) shows the event signature generated by the physical configuration of the MECL-RFID when the container is open. (b) When the lid is attached, the MECL-RFID event signature changes allowing for dynamic event notifications.**

## 5. Evaluating MPM and PRA Performance

We evaluate PRA performance in this paper via recognition error rate, which is defined as the percentage rate at which the expected result cannot be correctly recognized.

The MPM estimates complex pole patterns of the received signal from a chipless RFID tag, while the PRA recognizes the patterns of frequency notches of the signal. The estimated pole patterns or classified frequency-notch patterns can be used for detecting events by monitoring changes in event signatures. Note that the phase information provided through our network of CR RFID readers is not as accurate as the values obtained from an expensive vector network analyzer. To evaluate the performance of MPM and PRA, we consider both the cases when (*i*) there is no phase noise and (*ii*) phase noise exists. Evaluating the effects of chipless RFID tag decoding in the presence of phase noise is important to simulate real-world decoding, as the CR network used to scan MECL-RFIDs will introduce some phase noise internally at the receiver.

### 5.1 Decoder Performance with no Phase Noise

In our first simulation experiment, we evaluate chipless RFID error performance when Additive White Gaussian Noise (AWGN), without phase noise, is present for the MPM and the PRA decoding methods. An analytic signal using the SEM model with three poles (see Fig. 2 and Table 1) was used for the evaluation. Fig. 3 shows the error performance of the MPM and PRA as a function of the Signal-to-Noise Ratio (SNR) when there is only AWGN but no phase noise present. It is observed that if the SNR is higher than 18.5 dB, both methods show nearly zero error. Compared with MPM, the transition zone of the PRA-based approach is larger than MPM in terms of the error change with SNR.

### 5.2 Decoder Performance with Phase Noise

For the scenario where phase noise exists, the MPM's performance can be impacted significantly by phase noise. If one-degree uniform phase noise is added to the signal, as shown in Fig. 4, the normalized decoding error is more than 60%, even if the overall

**Table. 1. Complex pole parameters for SEM model**

| Index/Parameter | $\alpha$ | $\omega$ | R |
|---|---|---|---|
| 1 | $2\pi \times 10^8$ | $2\pi \times 10^9$ | 1.000 |
| 2 | $3.5\pi \times 10^8$ | $3.5\pi \times 10^9$ | 1.000 |
| 3 | $5\pi \times 10^8$ | $5\pi \times 10^9$ | 1.000 |

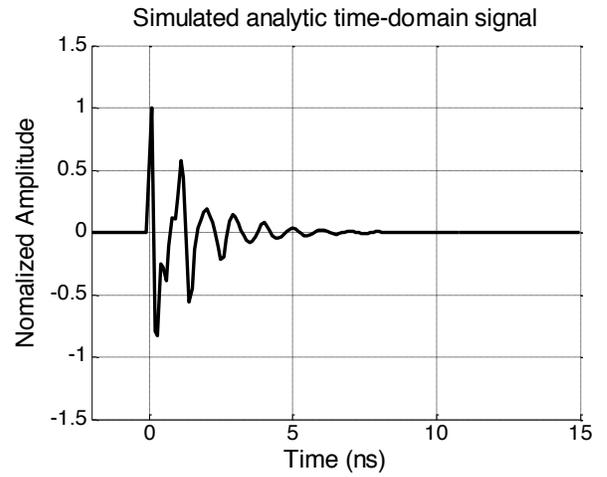

**Fig. 2. Simulated signal without any noise**

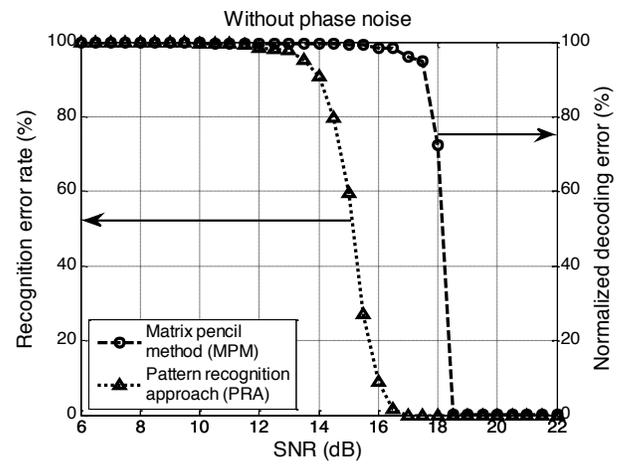

**Fig. 3. SNR versus error performance of MPM and PRA through simulations – AWGN only**

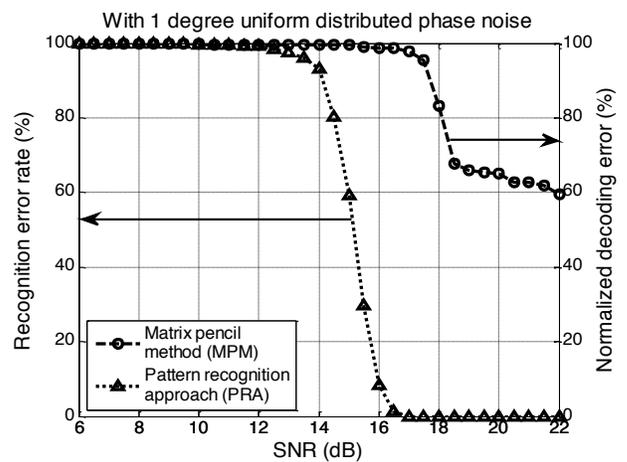

**Fig. 4. SNR versus error performance of MPM and PRA through simulations – AWGN and additional one-degree uniform phase noise**

SNR is high. In contrast, the PRA-based approach is not very sensitive to phase noise. This is because we only use the amplitude of the frequency domain signal for pattern recognition, while the MPM-based approach estimates the complex poles of the signal which requires both the amplitude information and the phase information.

Thus, compared with the MPM-based approach, PRA is better suited for implementation on a SDR testbed, which is much less expensive than a vector network analyzer but of lower performance in terms of calibration, sampling rate, and phase noise.

## 6. System Validation

### 6.1 Cyber-Physical Intervention Architecture

In earlier sections we introduced MECL-RFID (Section 4.2), which can be used to augment physical objects, such as medicine bottles, to report physical changes in the environment as they happen. We also discussed a CR network (Section 4.1) that can monitor changes in event signatures to ensure that drug events or any events of interest in the physical domain can be monitored in the cyber domain. Evaluating physical events in the cyber domain allows additional context to be combined with physical events to ensure relevant parties (e.g. patients, clinicians) are made aware of potentially hazardous ADEs. We now introduce the final piece in our proposed cyber-physical intervention system: the computing and message architecture in the cyber domain that monitors events and determines what appropriate action to take based on an event.

Figure 5 gives a high-level overview of the cyber and physical components which need to act in concert to monitor events in real-time, determine if any adverse events could potentially occur, and promptly inform clinicians or patients about events as they occur. The left side of the diagram shows how the cognitive radio network is constantly in a sensing loop, searching for changes in event signatures. The event sensing loop can be categorizes as follows:

1. Once a signature is read, the signature is sent to a computing node (which could be a service running in a medical cloud or a dedicated computer physically co-located on hospital premises), where post processing of the event signature is performed.

2. If the signal processing routine finds that a new event occurred, then the event is sent to an eMAR database and a

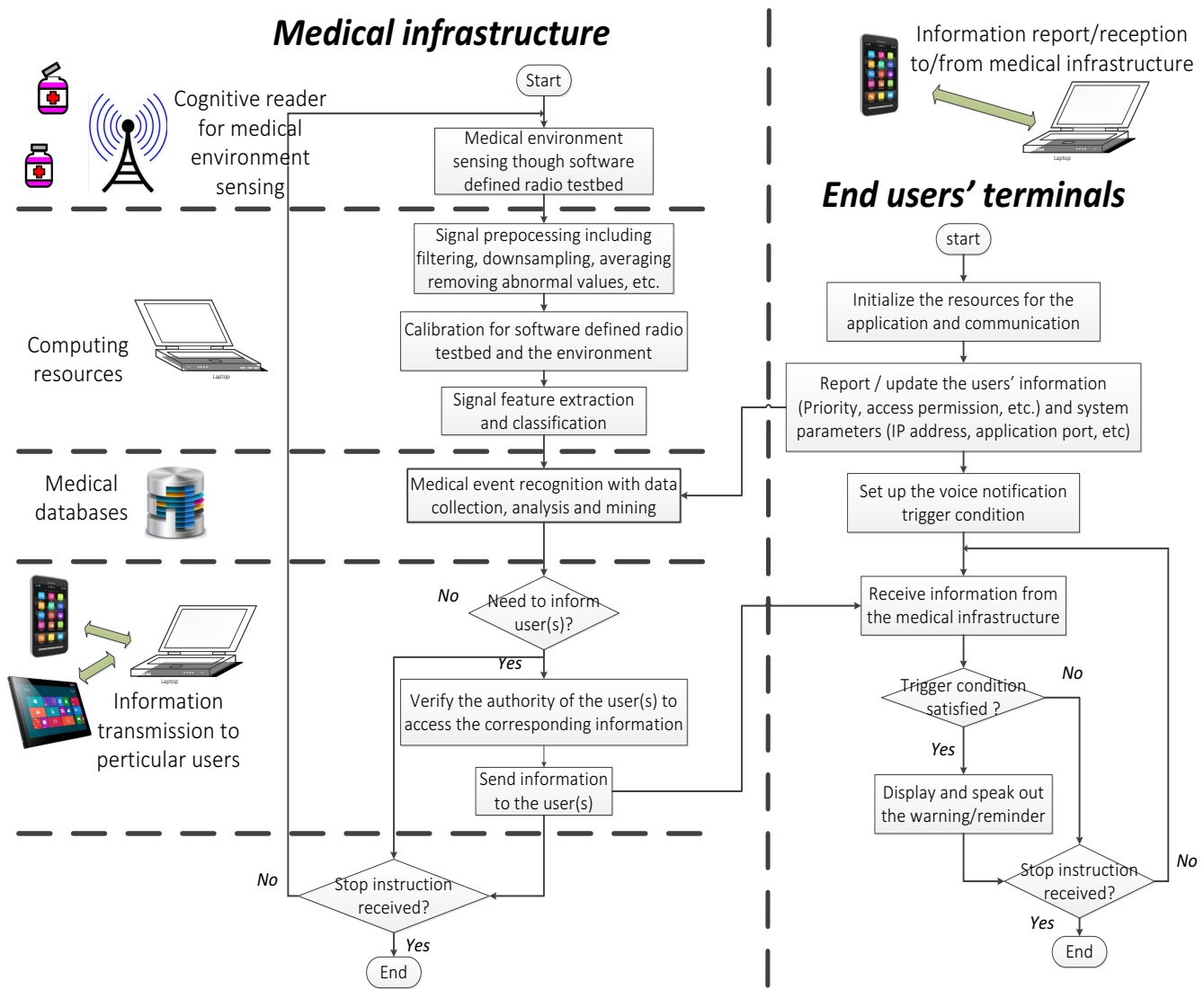

**Fig. 5. Cyber-physical intervention system components and high-level software flowchart.**

query is created to determine if any action needs to be taken based on the observed event signature.

3. The eMAR processes the query to determine which medicine bottle is reporting an event. The database then evaluates (*i*) the given medicine bottle and (*ii*) the patient the medication is about to be administered to. The eMAR determines whether potential ADEs could occur if medicine is administered.

4. The eMAR records the event.

5. If no intervention is required, then the listening loop starts again.

A. If the eMAR determines that an ADE could occur (e.g. one of the '5 rights' would be violated), then the eMAR sends a message to the patient or clinician about to administer the medicine with a warning message. This instant push message could be implemented using a protocol such as the eXtensible Messaging and Presence Protocol (XMPP) [21] that allows messages to be pushed to subscribed users.

B. The alert is delivered to the administering clinician's or patient's device where they are prompted to address the ADE before completing administration of the medication.

C. The clinician's or patient's response is recorded by the eMAR.

## 6.2 Research Prototype Implementation

Based on the cyber-physical intervention architecture and the event detection approaches, we created a stand-alone prototype system that encompasses all the system elements from Figure 5. The prototype system is shown in Fig. 6, where only the laptop we used to perform signal processing on the event signature is not shown. While we discuss utilizing multiple bits stored in the MECL-RFID tag, we implemented the minimal functionality to detect open and close notifications from an augmented medicine bottle. To demonstrate the capability of detecting the status of the medicine bottle (opened or closed), MECL-RFID tags that can store 1-bit information were designed and attached to the surface of a standard child-proof pill container and its lid (RFIDs #1 and #2 in Figure 6). A commercial software-defined radio, an Ettus Research USRP N210 [22], was used to simulate the cognitive radio network that excites the MECL-RFID and measures the amplitude of the received, scattered signals from the MECL-RFID in the pill container (i.e. measures the event signature).

To deal with the USRP N210's DC offset and noise present in the wireless environment, we implemented the MECL-RFID signal excitation and detection functionality as a SDR application in GNU Radio to filter the abnormal components and calibrate the SDR application to account for both the internal noise generated by the USRP and noise present in the wireless environment.

We implemented the PRA event detection algorithm (described in Section 4.3.2) on a Dell Latitude E6320 running Ubuntu 12.10. The event signatures we measured for open and closed-lid events are shown in Fig. 7. If the bottle is closed, the frequency notch can be found at around 1.6 GHz, while the frequency notch shifts to around 1.2 GHz when the bottle is opened. Based on the change of the resonance frequency caused by the tag coupling (i.e. the event signature is changed when the medicine bottle is closed, and the MECL-RFID's physical configuration is changed), the updated status of the medical container is detected by our USRP N210, and the time of the pill container event as well as the event that occurred is recorded. Although we chose to use a MECL-RFID operating in the 1.2 – 1.6 GHz band, MECL-RFID tags can be designed for use in other frequency bands, such as the FCC's proposed 3.5 GHz shared spectrum band, or industrial, scientific and medical (ISM) bands (e.g. 2.4 GHz or 5 GHz).

To evaluate real-time notification of detected event signatures, we developed a mobile application for Android to demonstrate the push notification service that would be provided by the eMAR. Any mobile device that subscribes to the eMAR service can receive the message as long as wireless connectivity is available (i.e. Cellular or WiFi). Based on the detection result of the PRA algorithm, our system can also instruct a mobile device to use an audible warning message. If the cyber-physical intervention system recognizes that the smart medical container was not opened during a pre-specified time, absence of an expected event can serve as a trigger for a push message that notifies the patient or the clinician that the expected medication administration may have been missed or should be taken shortly. Another case we investigated was double- or over-dosing: If the wireless cognitive radio network detects that multiple open and close event signatures occur within rapid succession, the patient or the clinician will receive the a push notification from the cyber-physical intervention system to prevent a potential double-dose ADE.

## 7. CONCLUSION AND FUTURE WORK

In this paper, we presented the concept of multi-element chipless RFID for augmenting medical containers as part of a cyber-physical intervention system that can seamlessly detect drug events such as a clinician opening a pill container in real-time. Simulation results indicated that our pattern recognition approach for decoding chipless RFIDs can detect chipless RFID tags more accurately than the MPM method in the presence of phase noise.

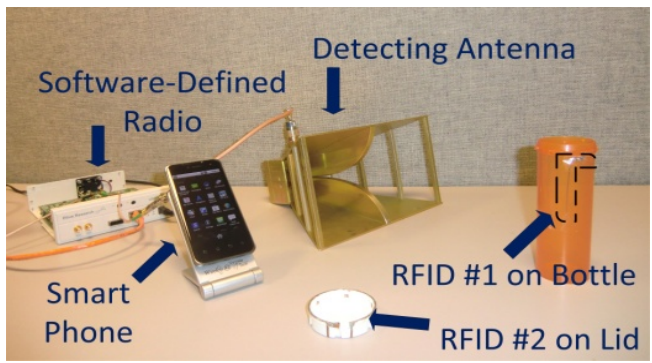

Fig. 6. Proof-of-concept prototype of smart medical container and event detection/notification system

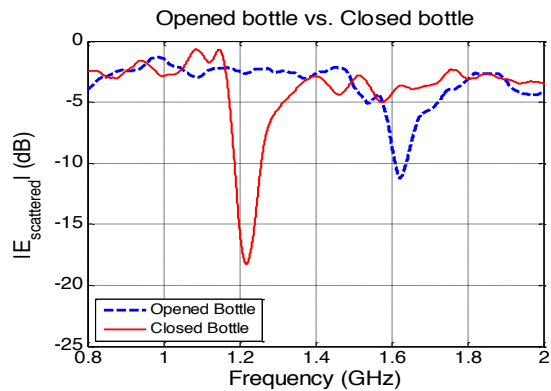

Fig. 7. Measured results of detected events including the closed bottle and opened bottle event signatures.

Our research prototype of the cyber-physical intervention system utilizing MECL-RFID and software-defined radio could effectively detect open and closed-lid events for an augmented pill container. The developed prototype system was able to monitor the series of events of the pill bottle, detect potential ADEs in real-time, and send a warning to a user through a smartphone. Furthermore, our prototype demonstrated that the PRA chipless RFID detection algorithm provided sufficient detection capabilities that could be evaluated in near real-time on consumer off the shelf (COTS) equipment.

Based on our research we feel further investigation on (*i*) counting pills and measuring liquid levels within a container, (*ii*) quantifying MECL-RFID tag detection range, (*iii*) performance of MECL-RFID in diverse and densely populated wireless environments, and (*iv*) clinical evaluations of MECL-RFID to detect events in an eMAR for healthcare workflows are needed as future work towards a MECL-RFID enabled workflow in a myriad of hospital and home care deployment scenarios.

## Acknowledgements

This work was funded in part by the Center for Advanced Engineering and Research (CAER), Virginia Tech's Institute for Critical Technology and Applied Science (ICTAS), and The Bradley Foundation Graduate Fellowship. We would like to thank Dr. Carl Dietrich for all of the helpful discussions throughout the course of this work, and our anonymous reviewers for their thoughtful and constructive feedback that allowed us to improve upon our original manuscript.